\documentclass[amsmath,amssymb, aps, reprint, prd,%preprintnumbers,%
nofootinbib, eqsecnum, superscriptaddress,showpacs,showkeys]{revtex4-1}
\usepackage{graphicx}% Include figure files
\usepackage{dcolumn}% Align table columns on decimal point
\usepackage{bm}% bold math
\usepackage{ulem}
\usepackage[T1]{fontenc} % if needed
\newcommand{\eqn}[1]{Eq.~(\ref{#1})}
\newcommand{\eqns}[2]{Eqs.~(\ref{#1}), (\ref{#2})}
\newcommand{\reference}[1]{Ref.~\cite{#1}}

\def\pa{{\partial}}

\usepackage{bbm}                                                  
\def\nc{\newcommand}
\nc{\beq}{\begin{equation}}  \nc{\eeq}{\end{equation}}
\nc{\bea}{\begin{eqnarray}}  \nc{\eea}{\end{eqnarray}}
\nc{\bal}{\begin{align}}  \nc{\eal}{\end{align}}
\nc{\baa}{\begin{array}}     \nc{\eaa}{\end{array}}
\nc{\bit}{\begin{itemize}}   \nc{\eit}{\end{itemize}}
\nc{\ben}{\begin{enumerate}} \nc{\een}{\end{enumerate}}
\nc{\bce}{\begin{center}}    \nc{\ece}{\end{center}}
\nc{\bpm}{\begin{pmatrix}}   \nc{\epm}{\end{pmatrix}}
\nc{\bvt}{\begin{verbatim}}  \nc{\evt}{\end{verbatim}}
\def\half{\frac12}	%\def\half{{1\over2}}

\def\to{\rightarrow}

\def\gesim{\gtrsim}
\def\lesim{\lesssim}
\def\gesim{\,{\raise-3pt\hbox{$\sim$}}\!\!\!\!\!{\raise2pt\hbox{$>$}}\,}
\def\lesim{\,{\raise-3pt\hbox{$\sim$}}\!\!\!\!\!{\raise2pt\hbox{$<$}}\,}
\def\boldoverdot{\,{\raise6pt\hbox{\bf.}\!\!\!\!\>}}

\def\ie{{\it i.e.}}
\def\eg{{$e.\,g.$}}

\def\bcal{{\cal B}}
\def\ccal{{\cal C}}
\def\dcal{{\cal D}}

\def\lcal{{\cal L}}
\def\mcal{{\cal M}}

\usepackage{bm}
\def\diag{\hbox{\diag}}

\def\kl{{\kappa\lambda}}
\def\mn{{\mu\nu}}

\def\doubleundertext#1{
{\undertext{\vphantom{y}#1}}\par\nobreak\vskip-\the\baselineskip\vskip4pt%
\undertext{\hbox to 2in{}}}
\def\inbox#1{\vbox{\hrule\hbox{\vrule\kern5pt
     \vbox{\kern5pt#1\kern5pt}\kern5pt\vrule}\hrule}}
\def\sqr#1#2{{\vcenter{\hrule height.#2pt
      \hbox{\vrule width.#2pt height#1pt \kern#1pt
         \vrule width.#2pt}
      \hrule height.#2pt} } }

\def\today{\ifcase\month\or
  January\or February\or March\or April\or May\or June\or
  July\or August\or September\or October\or November\or December\fi
  \space\number\day, \number\year}
\def\pmb#1{\setbox0=\hbox{#1}%
  \kern-.025em\copy0\kern-\wd0
  \kern.05em\copy0\kern-\wd0
  \kern-.025em\raise.0433em\box0 }

\def\pmbb#1{\setbox0=\hbox{#1}%
  \kern-.02em\copy0\kern-\wd0
  \kern.04em\copy0\kern-\wd0
  \kern-.02em\raise.03464em\box0 }
\def\sumprime_#1{\setbox0=\hbox{$\scriptstyle{#1}$}
  \setbox2=\hbox{$\displaystyle{\sum}$}
  \setbox4=\hbox{${}'\mathsurround=0pt$}
  \dimen0=.5\wd0 \advance\dimen0 by-.5\wd2
  \ifdim\dimen0>0pt
  \ifdim\dimen0>\wd4 \kern\wd4 \else\kern\dimen0\fi\fi
\mathop{{\sum}'}_{\kern-\wd4 #1}}
\def\viz{{\it viz.}}

\begin{document}
%\preprint{  {\bf Draft \jobname}, \today}
\preprint{LTH 1031, NSF-KITP-14-228, MCTP-14-45}

%---------------------------------------------------------------------------
\title{The Gauss-Bonnet Coupling Constant in\\
 Classically Scale-Invariant Gravity}

\author{Martin B Einhorn}
\email[]{meinhorn@umich.edu}
\affiliation{Kavli Institute for Theoretical Physics,\\
University of California, Santa Barbara CA 93106-4030,USA}\thanks{Current address}
\altaffiliation[also at ]{Michigan Center for Theoretical Physics,
University of Michigan, Ann Arbor, MI 48109-1040,~USA.}

\author{D R Timothy Jones}
\email[]{drtj@liv.ac.uk}
\affiliation{Kavli Institute for Theoretical Physics,\\
University of California, Santa Barbara CA 93106-4030,USA}
\affiliation{Department of Mathematical Sciences,\\ 
University of Liverpool, Liverpool L69 3BX,UK}\thanks{Current address.}

\date{\today}

\begin{abstract}
We discuss the renormalization of higher-derivative gravity, both without and with matter fields, in terms of two primary coupling constants rather than three.  A technique for determining the dependence of the Gauss-Bonnet coupling constant on the remaining couplings is explained, and consistency with the local form of the Gauss-Bonnet relation in four dimensions is demonstrated to all orders in perturbation theory.  A similar argument is outlined for the Hirzebruch signature and its coupling. We speculate upon the potential implications of instantons on the associated nonperturbative coupling constants.
\end{abstract}
%\arxivnumber{1412.5572}
\maketitle

%\tableofcontents
%---------------------------------------------------------------------------

\clearpage
%%%%%%%%
\section{Introduction}\label{sec:intro}
%%%%%%%%

The general form of the classically scale-invariant theory of the metric takes the form 
\begin{subequations}
\begin{align} \label{eq:action1}
S_1&\equiv \int_\mcal\!\!\! d^4x \sqrt{g}\, \lcal_1,\\ 
\label{eq:l1}
\lcal_1&\equiv\frac{1}{2\alpha}C^2+ \frac{1}{3\beta}R^2+\frac{2}{\gamma}\widehat{R}_\mn^{2} ,
\end{align}
\end{subequations}
where $C^2\equiv C_{\kl\mn} C^{\kl\mn}$ is the square of the Weyl
tensor,  $\widehat{R}_\mn\equiv R_\mn- g_\mn R/4$  is the traceless part
of the Ricci tensor, and $R=g^\mn R_\mn$  is the Ricci scalar.  The
integral is over the spacetime manifold $\mcal.$  (See
Appendix~\ref{sec:curvature} for the definition of the Weyl tensor). 
In general, the action\footnote{To be complete, a term $\Box R$ should be added as well, 
but it plays no role in the following. Like $G$, it appears in the conformal anomaly, but 
unlike $G$, it is the covariant divergence of a true vector $\nabla^\mu R.$}\eqn{eq:action1}
must be augmented by the addition of 
certain surface or boundary terms in
order to have the proper relationship for the composition of the path
integral~\cite{Gibbons:1976ue}. For the most part, such complications
will not concern us  since our interest is in perturbation theory, and
we shall ignore such terms  or simply assume that the manifold has no
boundaries. We shall adopt a Euclidean metric throughout this paper,
assuming that any spacetime in which we are interested can be realized,
with some choice of coordinates, through the replacement of one
coordinate, say, $x^4$ by $-ix^0,$ with a corresponding redefinition of
the metric $g_\mn$. 

Classically, the three quadratic invariants in \eqn{eq:l1} are in a sense not independent, because of the Gauss-Bonnet (G-B) relation, whose local form may be written as 
\begin{subequations}
\label{eq:gb}
\begin{align}
\label{eq:gbG}
G&\equiv C^2-2W,\ {\mathrm{where}}\ W\equiv \widehat{R}_\mn^2-\frac{R^2}{12},\hskip10mm\\
\label{eq:gb4}
G&\!=\! R^*R^*\!=\!R^*_{\kl\mn} R^*{}^{\kl\mn}, 
R^*_{\kl\mn}\! \equiv\!\textstyle{\frac{1}{2}}\epsilon_{\kl\rho\sigma}R^{\rho\sigma}{}_\mn.\!
\end{align}
\end{subequations}
$R^*_{\kl\mn}$ is the dual of the Riemann tensor. 
The first equation, \eqn{eq:gbG}, may be taken as the definition of $G$ in any dimension, whereas the second equation, \eqn{eq:gb}, is valid only in four-dimensions where the totally antisymmetric tensor is well-defined.  The fundamental result in four dimensions is that $G$ may be written locally as a 
divergence $G=\nabla_\mu B^\mu$ of a ``vector'' $B^\mu$ where 
\beq\label{eq:divb}
B^\mu\equiv
\epsilon^{\mn\gamma\delta}\epsilon_{\rho\sigma}{}^\kl\Gamma^\rho_{\kappa\nu}\left[ \half R^\sigma{}_{\lambda\gamma\delta}
+\frac{1}{3}\Gamma^\sigma_{\tau\gamma}\Gamma^\tau_{\lambda\delta} \right]\!,
\eeq
where $\Gamma^\kappa_\mn$ is the Levi-Civita connection associated with the metric.
In fact, $B^\mu$ does not transform as a vector under general coordinate transformations 
but transforms like a connection.  

Assuming that the four-manifold $\mcal$ is compact and has no boundaries,  
\beq\label{eq:intgb}
\int_\mcal\!\!\! d^4x \sqrt{g}\,G=32\pi^2\chi(\mcal),
\eeq
where the integer $\chi$ is the Euler characteristic of the manifold. 
If we rewrite the original Lagrangian\footnote{In \eqn{eq:l2}, we have used a different notation than in our earlier work, 
\reference{Einhorn:2014gfa}, for the coefficient of G, 
where it was called $\varepsilon$.}, 
\eqn{eq:l1}, as~\cite{Avramidi:1985ki, Avramidi:2000pia, Buchbinder:1992rb, Einhorn:2014gfa}
\beq\label{eq:l2}
\lcal_2\equiv \frac{1}{2a}C^2+\frac{1}{3b}R^2+ c\, G,
\eeq
then since $\sqrt{g}\,G$ is a total derivative, it makes no contribution to the equations of motion (EoM) and can be ignored, thereby reducing the classical theory from three parameters $(\alpha,\beta, \gamma)$ to two $(a, b)$.  
This is a bit glib, since, in a spacetime that is not Asymptotically Locally Euclidean (ALE), 
$G$ can give
a finite contribution to the classical action, even though it still
would contribute nothing to the EoM.

For future reference, a  commonly used Lagrangian~\cite{Stelle:1976gc, Fradkin:1981hx,*Fradkin:1981iu}, equivalent to 
\eqn{eq:l2}, utilizes the combination $W$ in \eqn{eq:gbG} instead of $C^2$, 
rewriting the Lagrangian density as 
\beq\label{eq:l3}
\lcal_3\equiv  \frac{1}{a}W+\frac{1}{3b} R^2+\widetilde{c}\, G.
\eeq
As a starting Lagrangian, one may choose $G$ together with any two other
linear combinations of $C^2, R^2,$ and  $\widehat{R}_\mn^2,$ so long as
they are linearly independent of $G$. One could not, for example, choose
$C^2,$ $W,$  and $G$~\footnote{Such a choice would correspond to  what
has been called conformal or Weyl gravity, with a Lagrangian involving
$C^2$ and $G$ (or $W$ and $G$). Such models presume that there is
a renormalization scheme free of the conformal anomaly.  No 
such a construction has never been displayed.  In
this paper, we assume that the anomaly exists and only consider models
renormalizable in that context.
}. Any such Lagrangian can be brought
to the form of \eqn{eq:l2}.  In that sense, the theory is unique.
 
What about the quantum field theory (QFT)?  A distinguishing property is
that the theory is renormalizable~\cite{Stelle:1976gc}, at least in a
topologically trivial background.  Since the topology ought not affect
the short-distance behavior of correlation functions, it is believed to
be renormalizable generally. Insofar  as perturbation theory is
concerned, the preceding three Lagrangian densities, when expressed in
terms of renormalized couplings and operators, require the addition of
divergent counterterms in order to obtain finite matrix elements as
functions of the renormalized coupling constants. In the process, the
operator $G$ cannot be ignored
because divergences arise that are not of the form of 
linear combinations of $W$ or $R^2$, 
but require a third invariant~\cite{Fradkin:1981hx,
*Fradkin:1981iu}.
 
A fundamental difference between the classical theory and the QFT is
that the latter is not scale-invariant after renormalization.  In the
context of such a scale or conformal anomaly, one might well wonder
whether the G-B relation is also anomalous~\cite{Brunini:1993qz}.  On
the other hand, the G-B relation, especially in its integral form, is a
generic result in topology\footnote{For an introduction to differential
geometric concepts, see, \eg, \reference{Eguchi:1980jx,
*AlvarezGaume:1984dr, *AlvarezGaume:1983at, Alvarez:1988tb}.  In its most
general form, it does not even require a metric~\cite{Chern:1944}.}. 
Like the Bianchi identities, to which it is related, it would be
disturbing if the four-dimensional QFT did not recover these
topologically-based identities. 

 Assuming that the G-B relation holds in four dimensions in the QFT, then, under any small variation of the metric $g_\mn \to g_\mn+\delta g_\mn$, the variation of the action is zero, 
\beq\label{eq:varindex}
\delta \int\!\!d^4x\,\left(\sqrt{g}\, G\right)=
\int\!\!d^4x\,\pa_\mu\,\delta\!\left(\!\sqrt{g}\,B^\mu\right)=0.
\eeq
By ``small variation,'' we mean any variation that does not change the
spacetime topology.  Although it is peculiar to four dimensions, this
relation is an algebraic identity and does not depend on any assumptions
about the background or require any reference to the EoM. As a result,
researchers have tended to ignore $G$ when formulating the Feynman rules
for this theory, even though it is essential for renormalizability.  

One source of confusion is that the preferred form of gauge-invariant
regularization, \viz, dimensional-regularization (DREG) requires that we
entertain the meaning of the theory outside of four-dimensions.  While
there are alternative possibilities for a four-dimensional,
gauge-invariant regularization, such as the generalized zeta-function
method, it is not so clear that they are implementable beyond one-loop. 
In any case, unlike the dual operators in \eqn{eq:gb4}, it  possible to
generalize \eqn{eq:gbG} to any dimension, so one can expect to recover
this linear combination when returning to four dimensions.

In this paper, we wish to make explicit that the renormalized theory can
be made consistent with the G-B relation and derive a relation between
the renormalization of $c$ and the renormalizations of the other two
couplings, say, $a,b$.  One may use DREG, and it is not even necessary
to modify the usual mass-independent renormalization procedures such as 
minimal subtraction (MS).  It is necessary to reinterpret the way in
which the reduction from three couplings to two has been achieved by
previous authors, especially since it plays an important role in our 
earlier discussion of dimensional transmutation~\cite{Einhorn:2014gfa}. 
Since the present paper is the companion promised there,
\reference{Einhorn:2014gfa} will henceforth be referred to as [{\bf I}].

We conclude this introduction with an outline of the remainder of the
paper.  In the next section~\ref{sec:puregrav}, we discuss aspects of
perturbative renormalization for the higher-order theory without matter.
 This is divided into three parts, reviewing the rather confusing
history and status of this puzzle, the details of renormalization using
DREG and MS, and finally some insights that may be gleaned by use of the
renormalization group equations (RGE).  Then, in
Section~\ref{sec:matter}, we indicate how our results may easily be
extended to include matter fields.  In Section~\ref{sec:hirzebruch}, we
discuss briefly another topologically significant parameter, the
Hirzebruch signature, that will enter discussions of the axial anomaly,
CP-violation, and related issues, such as the $U(1)$-problem in QCD. 
That leads us to speculate, in Section~\ref{sec:instantons}, about the
potential role that gravitational instantons, a nonperturbative effect, 
may have on some of these considerations.  Finally, in
Section~\ref{sec:conclusions}, we end with a summary of results and some
important remaining questions.  Two appendices have been added to
clarify some issues in background field quantization
(Appendix~\ref{sec:bfq}) and in the extension of curvature to
$n$-dimensions (Appendix~\ref{sec:curvature}).

\section{Perturbative Renormalization--Pure Gravity}\label{sec:puregrav}

\subsection{History and Framework}\label{sec:history}

%\paragraph{ }
  
In their seminal papers on this theory, Fradkin \&
Tseytlin~\cite{Fradkin:1981hx,*Fradkin:1981iu} adopted the form
\eqn{eq:l3}.  They initially state that the topological term $G$ can be
``disregarded'' under the usual assumptions, such as the ``natural
asymptotically flat boundary conditions.''  Nevertheless, after
obtaining the Feynman rules, which of course requires the addition of
gauge-fixing terms and Faddeev-Popov ghosts, they find that there are
gauge-invariant divergences not only of the tensor structure of the
operators $W$ and $R^2$, but also of the form of $G$.  They therefore
assign a counterterm to  $c\, G$, which, because the Feynman rules are
independent of  $c,$ depend only on the other parameters of the theory. 
That is, the counterterms assigned to the ``coupling constant'' $c$ are
independent of $c$.  However, when one goes beyond one-loop order, one
might think that one must include vertices involving such counterterms
for $G$ in addition to those for $W$ and $R^2.$  Although they feel no
need to modify their Feynman rules, this is an unusual prescription, and
it is unclear what is precisely going on.  In particular, it is not so
clear that, when $G$ is expressed as a linear combination of the three
renormalized operators as in \eqn{eq:gb}, the resulting renormalized
Lagrangian in four dimensions necessarily obeys \eqn{eq:varindex}.

Similarly, Avramidi \& Barvinsky~\cite{Avramidi:1985ki} and Buchbinder
{\it et al.}\/~\cite{Buchbinder:1987et} choose a Lagrangian density of
the form \eqn{eq:l2}.   Buchbinder {\it et al.}\/ state (below their
eq.~(8.3)), that the topological term can only make a finite
contribution to the one-loop corrections and, for k-loops, will only
contribute to the poles in $1/(n-4)^{k-i}$ with $i\ge1.$  Nevertheless,
in their elaboration of the one-loop divergences (see their
eqs.~(8.102),(8.103)), they encounter a divergent, one-loop contribution
to the term $c\, G$.  In fact, in \reference{Buchbinder:1987et}, it is
stated and assumed that the action  without the $G$ term is
multiplicatively renormalized, which is not true. These paradoxes derive
from the conflict between using DREG, on the one hand, and a
four-dimensional identity \eqn{eq:varindex} on the 
other\footnote{Some of these considerations were taken up in
\reference{deBerredoPeixoto:2004if}, which also considered the nature of
the theory for finite but small $\epsilon=4-n$.}.  Beyond one-loop
order, it is not obvious that this conflict can always be resolved.  

We submit that a consistent formulation exists that regards the beta-function 
$\beta_c$ as determined by the beta-functions of the other coupling constants. The point is that, as in the procedure adopted in \reference{Fradkin:1981iu} at one-loop, the counterterms for $c$ are determined by the counterterms for the other couplings, $a,b$.  
This suggests that we regard $c = c(a,b)$, a function of the other couplings, satisfying the consistency relation
\beq\label{eq:consistency}
\frac{\pa c}{\pa a}\beta_a
+\frac{\pa c}{\pa b}\beta_b=\beta_{c}.
\eeq
 We shall shortly prove this, viz., $c$ is indeed a function of $a,b$ that 
obeys \eqn{eq:consistency}. This equation will be shown to determine 
the function $c$ up to its initial value $c_0.$

In fact, the $\beta$-function $\beta_c$ described above 
represents a generalisation  
to the quantised $R^2$-gravity case of the Euler anomaly
coefficient, and thus a candidate for an 
$a$-function as proposed by Cardy~\cite{Cardy:1988cwa}, manifesting a
4-dimensional $c$-theorem.  Results for this anomaly coefficient
(without quantising gravity) include  a non-zero 5-loop contribution
involving four  quartic scalar couplings~\cite{Hathrell:1981zb}  and
non-zero three loop contributions involving gauge and Yukawa
couplings~\cite{Freedman:1998rd}. (For some recent progress  on the
$a$-theorem and references, see Refs.~\cite{Jack:2013sha,*Jack:2014pua}.)

The  relation $c=c(a,b)$ or \eqn{eq:consistency} is reminiscent of the
method of coupling constant reduction by Oehme and
Zimmermann\footnote{For reviews with references to earlier works, see
\reference{Zimmermann:1984sx, *Oehme:1985jy, *Oehme:2000eg,
*Oehme:1999fd, *Zimmermann:2001pq}.}, employed  to seek general
relations among renormalized coupling constants that were
renormalization group invariant. Their method leads to nontrivial
constraints on the parameters of the theory, whereas, in the present
case, the relation is a direct consequence of the renormalization
properties of the theory. This case is similar in the following sense: 
Suppose that you had started from \eqn{eq:l1} with three coupling
constants $\alpha,\beta,\gamma.$ Each of the three operators can be
defined in $n$-dimensions. (See Appendix~\ref{sec:curvature}.) So you
can use DREG to regularize and MS to renormalize this theory consistent
with gauge invariance.  Then, having obtained a finite renormalized
theory in four-dimensions, you might ask whether there is some relation
among the three renormalized couplings and eventually discover that, for
certain linear combinations of couplings, only two linear combinations
appear in the beta-functions. So you might eventually arrange them in
the form of, say, \eqn{eq:l2} or \eqn{eq:l3}, hypothesize that
$c=c(a,b),$ and discover that the relation \eqn{eq:consistency} can be
imposed, in effect, reducing the number of couplings from three to two.

On the other hand, the present situation is dissimilar from coupling
constant reduction since, in order to recover the Bianchi identities and
maintain the G-B relation, properties that the theory in four-dimensions
must have, the relations among the couplings are essential. These
relations act like additional symmetries, but ones that only hold in
four dimensions. They cannot be anomalous since their validity makes no
reference to the EoM or to a conserved current resulting from a
symmetry. They are constraints that must follow for a sensible
gauge-invariant, renormalized theory in four-dimensions, not a
hypothesis to be tested. 

To show that \eqn{eq:consistency} is satisfied, it may be helpful to
define  $w\equiv a/b$ and to rewrite the Lagrangian density,
\eqn{eq:l2}, as\footnote{As explained in I, there are reasons why 
it would be more logical to use the ratio $b/a$ rather than $a/b$, but, as before, 
we choose to remain faithful to the usual convention.}   

\beq\label{eq:l4} 
\lcal_4\equiv\frac{1}{a}\left[\half C^2+\frac{w}{3}R^2\right]+c\, G.
\eeq
We imagine quantizing the theory by the background field method, as briefly explained in Appendix~\ref{sec:bfq}.  We presume that gauge-fixing is done in a manner consistent with background field gauge invariance.  In fact, we shall suppress gauge-fixing parameters and ghost terms in the following, because our concern will be with the gauge-invariant beta-functions.  Assuming \eqn{eq:varindex}, we can ignore 
$c\, G$ in formulating our Feynman rules. Then we may identify $a$ with the loop-counting parameter.
 
Although a slight digression, a word of warning must be added.  Using a running coupling to count loops only works so long as the renormalization scale is held fixed.  Recall from~[{\bf I}] that, at one-loop order, the scale dependence of $a$ is 
\beq\label{eq:leadinglogs}
a(\mu)=\frac{a_0}{1+a_0L_t}=a_0-a_0^2L_t+a_0^3L_t^2+\ldots...,
\eeq
where $L_t\equiv\kappa\beta_2 t,$ with $t\equiv\ln(\mu/\mu_0);$  
$\kappa$ and $\beta_2$ are constants. Thus, the coupling constant at
scale $\mu$ involves the coupling constant $a_0$ at some reference scale
$\mu_0$ to arbitrary orders in $a_0.$   This observation becomes
especially important in higher loops or, even at one-loop, for couplings
such as $w$ that mix with others.  To discuss the renormalization group,
one must use a different, fixed parameter such as $\hbar$ to count
loops.  These  seemingly trivial observations will become extremely
important below when a function of $w(\mu)$ will be re-expressed in
terms of $a(\mu),$ at the same order in the loop
expansion. (See \eqn{eq:rgiea} below.)

\subsection{Renormalization in Detail}\label{sec:renorm}

First, we shall review some details of DREG and MS to establish notation
and to emphasize certain features of MS.  Following
't~Hooft~\cite{'tHooft:1973mm}, we renormalize the couplings $a$ and $w$
as follows:
\beq\label{eq:arenorm}
\frac{1}{a_B}=\mu^{-\epsilon}
\left[\frac{1}{a}+\frac{A_1(a,w)}{\epsilon}\!+\!\frac{A_2(a,w)}{\epsilon^2}+\!\ldots\right],
\eeq
where $n\equiv 4-\epsilon$, $\mu$ is the renormalisation scale, 
and the ellipses represent higher order terms in powers of $1/\epsilon$.  
The factor $\mu^{-\epsilon}$ in front appears in order to make the renormalized coupling $a$ dimensionless, independent of the dimension $n$.
Similarly,
\beq\label{eq:warenorm}
\frac{1}{b_B}\!\equiv\!\frac{w_B}{a_B}\!=\!\mu^{-\epsilon}\!\left[\frac{w}{a}\!+\!\frac{B_1(a,w)}{\epsilon}
\!+\!\frac{B_2(a,w)}{\epsilon^2}+\ldots\right]\!,\!
\eeq
or, dividing by \eqn{eq:arenorm},
\beq\label{eq:wrenorm}
w_B=w+\frac{a(B_1-wA_1)}{\epsilon}+\ldots.
\eeq
From \eqn{eq:arenorm}, the variation of $a$ with scale $t\equiv\ln(\mu/\mu_0)$ 
is given by
\beq
0\!=\!-\!\epsilon\left[\frac{1}{a}\!+\!\frac{A_1(a,w)}{\epsilon}\right]\!+
\frac{da}{dt}\left[-\frac{1}{a^2}\!+\!\frac{1}{\epsilon}\frac{\pa A_1}{\pa a}\right]\!+
\frac{dw}{dt}\frac{1}{\epsilon}\frac{\pa A_1}{\pa w}+\!\ldots .\!\\
\eeq
To obtain the beta-functions, we want to isolate the terms of $O(1)$ or higher in $\epsilon.$  
(In order for the theory to be renormalizable, terms involving negative powers of $\epsilon$ 
must cancel among themselves~\cite{'tHooft:1973mm} in the limit $\epsilon\to0.$)
As expected from its definition, the coupling $w_B$ is dimensionless for all $n$, 
so $dw/dt$ will have no terms of order $\epsilon$, and the last term can be neglected.  
Then we find
\beq\label{eq:ba}
\frac{da}{dt}=-\epsilon a + \beta_a,\ {\rm{with}}\ \beta_a=-a^2\frac{\pa (aA_1)}{\pa a}.
\eeq
Similarly, from \eqn{eq:wrenorm},
\beq\label{eq:bw}
\frac{dw}{dt}=\beta_w=a\frac{\pa }{\pa a}\left[a(B_1-wA_1)\right].
\eeq
The counterterms $A_n, B_n$ may in principle be calculated
order-by-order  in the loop expansion.  In a given order $N$, the
counterterms $A_n,B_n$ vanish for $n\ge N+1,$ so there are only a finite
number of counterterms to each order.  Further, as 't~Hooft 
showed~\cite{'tHooft:1973mm}, at a given order, the counterterms
$A_n, B_n$ for  $n\ge 2$ are completely determined the results of
lower-order calculations.  (This is why the beta-functions depended only
on $A_1$ and $B_1$ to each order.)  We exploited this fact in [{\bf I}] to
determine the dilaton mass, which first arises at two-loops in this
model, from the results at one-loop.

The one-loop divergences have been calculated~\cite{Fradkin:1981hx,*Fradkin:1981iu,
Avramidi:1985ki} with the result that
\beq\label{eq:one-loopct}
A_1=\beta_2=\frac{133}{10},\ B_1=\beta_3(w)=\frac{10w^2}{3}-5w+\frac{5}{12}.
\eeq
Thinking for a moment of $a$ as a loop-counting parameter, with the tree
approximation of order $1/a,$ it comes as no surprise that the one-loop
divergences are independent of $a$.
This is why, in [{\bf I}], we found that at one-loop, $\beta_w = a\overline{\beta}_w(w)$, with 
$\overline{\beta}_w(w)=\beta_3(w)-w\beta_2.$ (As we shall see in Section~\ref{sec:rge}, 
this scaling relation will not persist in higher orders.) 

Presuming that these beta-functions are known, at least to some loop
order, we wish to obtain the running couplings $a(t), w(t)$ by solving
the coupled system of equations,
\beq
\frac{da}{dt}=\beta_a(a,w),\quad \frac{dw}{dt}=\beta_w(a,w), 
\eeq
where we have taken the limit $\epsilon \to 0$ in \eqn{eq:ba}.
It is well known that the general solution of a first-order system of
this kind is unique up to the specification of the initial values
$(a_0,w_0)$ at some reference scale $\mu_0,$ which we have defined to be
$t=0.$  The fixed points of the system are obtained from the
simultaneous zeros of the beta-functions $\beta_a(a,w)=0,
\beta_w(a,w)=0.$  As remarked above, at one-loop order, 
$\beta_a=-\beta_2 a^2$ for a positive constant $\beta_2.$  
Within the perturbative regime, we may conclude 
that $\beta_a/a^2<0$ to all orders, so that $a(t)$ is monotonically decreasing from its initial value $a_0>0.$  Thus, in this simple model, the fixed points are  determined by the zeros of $\beta_w.$ 

Since the counterterms $A_1$ and $B_1$ can in principle be calculated
order-by-order in the loop-expansion, the beta-functions
$\{\beta_a(a,w), \beta_w(a,w)\}$ can be presumed known to arbitrary
order.
The running couplings  $a(t),w(t)$ are therefore in
principle known from the solutions to their defining equations,
\eqns{eq:ba}{eq:bw}, up to their initial values $a_0,w_0.$ We now want
to discuss the coupling $c$ and its beta-function. As described earlier,
having chosen Feynman rules that are independent of the coupling $c$,
its counterterms,  $C_n(a,w)$ are also completely fixed in terms of
$(a,w).$  For example, the counterterm  $C_1(a,w)/\epsilon$ is
determined by what is ``left over'' from the divergences assigned to
$A_1(a,w)/\epsilon$ and $B_1(a,w)/\epsilon$ (as well as any contribution
to $\Box R,$ which, as discussed earlier,  we can ignore.) The
renormalization of $c$ therefore proceeds more or less like the
renormalization of any other coupling constant, 

\beq\label{eq:cbare}
c_B = \mu^{-\epsilon}\left[c(\epsilon)+\frac{C_1(a,w)}{\epsilon}+\frac{C_2(a,w)}{\epsilon^2}+\ldots\right],
\eeq
where we assume that the function $c(\epsilon)$ may be expanded as a
power series in  $\epsilon$ with nonnegative powers, so that the
renormalized coupling  $c\equiv\lim_{\epsilon\to0} c(\epsilon)$ exists.
What is different about the renormalization of $c$ is that all the
counterterms $C_n(a,w)$  are independent of $c$.
Hence, we have\footnote{There are potentially negative powers of 
$\epsilon$ in \eqn{eq:bc} in addition to these nonnegative ones, but, as 't~Hooft~\cite{'tHooft:1973mm} showed, these all must cancel among themselves.}
\begin{subequations}
\begin{align}
\label{eq:bce}
\frac{dc(\epsilon)}{dt} &= \epsilon c(\epsilon) + \beta_c,\\
\label{eq:bc}
\mathrm{where}\ 
\beta_c(a,w) &=  \frac{\pa (aC_1(a,w))}{\pa a}
\end{align}
\end{subequations}
The one-loop calculation~\cite{Fradkin:1981hx,*Fradkin:1981iu,Avramidi:1985ki} gives    
$C_1=-\beta_1$ with $\beta_1=+196/45,$ a constant.

Given its defining equation 
(now taking the limit $\epsilon \to 0$ in \eqn{eq:bce}), 
\beq\label{eq:betac}
\frac{dc}{dt}=\beta_c(a(t),w(t)),
\eeq
with the running couplings $a(t),w(t)$ and the function $\beta_c(a,w)$ presumed known,
the formal solution is 
\beq\label{eq:ct}
c(t)-c_0=\int_0^t\!\! dt' \beta_c(a(t'),w(t')).
\eeq
Thus, the renormalized coupling $c(t)$ is completely determined up to its initial value $c_0,$ which was our first claim.  \eqn{eq:ct} does not determine that 
$c(t)-c_0$ is a function of $(a(t),w(t))$ at the same scale $t$; it appears to depend upon their history, \ie, it appears as if $c(t)-c_0$ is actually a functional $F[a(t),w(t)].$  That is an illusion.  
Since $a(t)$ is monotonically decreasing, its inverse $t=t(a)$ is well-defined, so that, in the integral in \eqn{eq:ct}, we can change variables from $t'$ to $a'$, writing
\beq\label{eq:cat}
c(t)-c_0=\int_{a_0}^{a(t)} \!\!\!da' \frac{\beta_c(a',w(t(a') ))}{\beta_a(a',w(t(a') ))}.
\eeq
This shows that $c(t)$ is actually an ordinary function of the value $a(t)$ at the same scale. Further, it shows that the only $t$ dependence of $c(t)$ is implicit through its dependence on $a(t),$ just like other couplings.  A similar argument applies to the dependence on $w.$ By definition, $\beta_w(a,w)$ does not vanish except at a fixed point, so that, within a given phase, $\beta_w(a,w)$ will have a definite sign.  Therefore, although $w(t)$ may be increasing or decreasing, it too is monotonic and may be inverted $t=t(w).$  As with $a(t),$ we may change variables in \eqn{eq:ct} from $t'$ to $w'$ to establish that $c(t)$ only depends on the function $w(t')$ through its value $w(t)$ at the same scale.  Therefore, we have established our second claim\footnote{This conclusion may be generalized to include additional dimensionless coupling constants $\lambda_i$ associated with the inclusion of matter, but, since not all couplings necessarily run monotonically, the preceding argument must be modified slightly.  As the couplings $\{w(t),\lambda_i(t)\}$ evolve, the interval $(0,t)$ may be broken up into a finite number of closed subintervals $[0,t_1],[t_1,t_2], \ldots, [t_N,t]$ between which all the couplings run monotonically. Since the couplings are continuous, they must agree at the end-points $t_p$.  Thus, the result may be built up piecewise.}: 
$c-c_0\!=\ccal(a(t),w(t))$ for some function $\ccal.$ 

To summarize what has been determined thus far, in \eqn{eq:ct}, we have displayed a solution to  
\eqn{eq:betac}.  If we know the functions $a(t),b(t),$ then the solution is unique up to the constant $c_0.$  Further, we know that the functions $a(t),b(t)$ are uniquely determined by the values of $(a_0,b_0).$  If we do not know these initial values, then we could regard the solution in \eqn{eq:ct} as a three-parameter family of solutions $c(t;a_0,w_0,c_0).$  

If this were a real theory of nature rather than a model, we believe that, in principle, $(a_0,w_0)$ would be experimental observables.  We have not determined that $c_0$ is observable, and we will return to this question later.  We know in addition that $c(t)$ is in fact a function of $(a,w)$, \ie, $c=c_0+\ccal(a(t),w(t)).$  Can we say more about the function $\ccal(a,w)$?
The answer is yes since $c$ obeys the renormalization group equations.

\subsection{Renormalization Group Equations} \label{sec:rge}

%\paragraph{}
Knowing that $c(t)-c_0$ is a function of $(a(t),w(t))$, we may write
\begin{subequations}
\label{eq:rge}
\begin{align}
\label{eq:cawt}
\frac{\pa c}{\pa a}\,\frac{da}{dt}
\,&+\, \frac{\pa c}{\pa w}\,\frac{dw}{dt}\,=\,\frac{dc}{dt},\\ 
\label{eq:irge}
\mathrm{or}\ \beta_a(a,\!w)\frac{\pa c}{\pa a}\!&+\! \beta_w(a,\!w)\frac{\pa c}{\pa w}\!=\!\beta_c(a,\!w).
\end{align}
\end{subequations}
These equations are very powerful; each is a form of the RGE for the function $\ccal(a,w).$ One of its applications is to relate the functions in different orders in perturbation theory.  For example, it is clear from \eqn{eq:irge} that the one-loop approximation to the three beta-functions constrains the tree-approximation to the function $\ccal(a,w).$  The two-loop approximation to the beta-functions will constrain the one-loop correction to $\ccal(a,w),$ etc.

Note that \eqn{eq:irge} makes no reference to the scale parameter $t$ and poses the problem of finding $c$  as one of determining the solutions of a first-order, inhomogeneous partial differential equation.  Although these equations are not linear in c, the difference between any two solutions satisfies the homogeneous equation, which is linear. The generic approach to the study of such equations employs the method of characteristics\footnote{See, \eg, \reference{Arnold:1978}. For an application in an analogous context, see \reference{Einhorn:1982pp}. If  the initial values $(a_0,w_0,c_0)$ are regarded as unknown, this method can provide insight into the manifold of all solutions.}.  In the present context, however, we believe that it is simpler to exploit the loop expansion, especially because nothing much is known beyond one-loop order about theories of this type.

We may take advantage of the fact that  \eqn{eq:irge} makes no explicit reference to the scale parameter to parameterize the loop expansion in terms of the coupling $a$, at some fixed scale.  In particular, the counterterms may be expanded as 
\begin{align}\label{eq:loopcounter}
\begin{split}
A_1&=\sum_{k=1}^\infty a_{k}(w)a^{k-1},\  B_1=\sum_{k=1}^\infty b_{k}(w)a^{k-1},\cr  
C_1&=\sum_{k=1}^\infty c_{k}(w)a^{k-1},
\end{split}
\end{align}
where the $a_k, b_k, c_k$ corresponds to the $k^{th}$ term in the loop-expansion. 
Then, from \eqns{eq:ba}{eq:bw} and \eqn{eq:bc}, we have
\begin{align}\label{eq:loopbeta}
\begin{split}
-\frac{\beta_a}{a^2}&=\!\sum_{k=1}^\infty k\,a_{k}(w)\,a^{k-1},\cr 
\frac{\beta_w}{a}&=\!\sum_{k=1}^\infty k\,w_k(w) a^{k-1}, 
\beta_c =\!\sum_{k=1}kc_{k}(w)a^{k-1},
\end{split}
\end{align}
where, for brevity, we defined $w_k(w)\!\equiv\! b_{k}(w)\! -\!w a_{k}(w).$
Similarly, we may expand $c(a,w)$
\beq\label{eq:loopcaw}
c(a,w)=c_0+\frac{e_0(w)}{a}+\sum_{k=1}^\infty e_k(w)a^{k-1},
\eeq
where, in addition to the constant $c_0,$ a tree-level contribution $e_0(w)/a$ has been included.

To determine $e_0(w)$ explicitly, we must insert the one-loop contributions to the beta-functions into \eqn{eq:irge} to obtain 
\begin{subequations} 
\label{eq:e0}
\begin{align}
\label{eq:e0a}
 a_1(w) e_{0}(w)+w_1(w)e_0'(w) &= c_1(w),\\
\label{eq:e0b}
 \beta_2 e_0(w)+\overline{\beta}_w(w) e_{0}'(w) &=-\beta_1,
\end{align}
\end{subequations}
where, in \eqn{eq:e0b}, we inserted the one-loop values for $a_1,c_1$ and 
$w_1(w)=\overline{\beta}_w(w)$ from from \eqn{eq:one-loopct} and from immediately below \eqn{eq:bc}.
 The actual values are not so important as the fact that $\beta_1,\beta_2$ are constants.  Then we observe that a solution of 
\eqn{eq:e0b} is simply $e_0(w)=-\beta_1/\beta_2,$ a constant, regardless of the form\footnote{In other words, we would not need to know the one-loop renormalization $b_3(w)$ of the $R^2$ term.  Although we will not demonstrate it here, this persists in higher orders in the sense that, in order to determine the $O(N\/)$-loop contribution to $c,$ we only need know the renormalization of $R^2$ to $O(N\!-\!1).$}
Therefore, the tree approximation to $c$ is 
\beq\label{eq:treec}
c_(a,w)=c_0-\frac{\beta_1}{\beta_2 a}.
\eeq
This is a rather remarkable result in some ways.  As advertised, the one-loop beta-functions in \eqn{eq:irge} determine the tree approximation for $c.$  On the other hand, unlike ordinary coupling constants, the only arbitrariness in $c$ is the constant $c_0,$ so rather than a consistency check, the RGE actually {\bf determines} the tree approximation.  Even though $\beta_1,\beta_2$ are quantum corrections of $O(\hbar),$ their ratio is $O(1).$ 

It is convenient but not crucial that the one-loop corrections $\beta_1,\beta_2$ be independent of $w$; however, if $\beta_2$ were dependent on $w$, there may be a danger that their ratio would either be singular or vanish for certain values of $w$. Nevertheless, there remains a paradox: 
although \eqn{eq:treec} corresponds to one solution, there appear to be others, since, to any solution $e_0(w)$ of \eqn{eq:e0b} may be added a solution $e_h(w)$ of the homogeneous equation
\beq\label{eq:hrge}
\beta_2 e_h(w)+\overline{\beta}_w(w)e_h'(w)=0.
\eeq
Thus, we could replace the solution \eqn{eq:treec} by
\beq\label{eq:treech}
c_(a,w)=c_0+\frac{e_h(w)}{a}-\frac{\beta_1}{\beta_2 a}.
\eeq
On the other hand, we argued earlier that the solution \eqn{eq:ct} was unique up to the constant $c_0.$  How can both statements be true?  The answer is that, like $c_0$, ${e_h(w(t))}/{a(t)}$ is renormalization group invariant, \ie, to one-loop order, it is independent of $t.$  This is easily seen:
\begin{align}\label{eq:rgiea}
\begin{split}
\frac{d}{dt}\left(\!\frac{e_h(w(t))}{a(t)}\!\right)
&\!=\!-\frac{\beta_a}{a^2}\,e_h(w)\!+\!e'_h(w)\frac{\beta_w(a,w)}{a}\!\cr
&=\beta_2\,e_h(w) +e'_h(w)\overline{\beta}_w,
\end{split}
\end{align}
which is identical to \eqn{eq:hrge} and therefore zero.  (Recall
our earlier warning surrounding \eqn{eq:leadinglogs}.)   Thus, the
ambiguity simply corresponds to the freedom to choose a different value
of $c_0.$  (It is easy enough to verify this explicitly by writing down the general 
solution of \eqn{eq:hrge}, using $\beta_2$ and $\beta_3 (w)$ as defined 
in \eqn{eq:one-loopct}). Our general arguments above assure us that this result
remains true to arbitrary order in perturbation theory; we can choose
any solution for the $e_k(w)$ and the ambiguity can eventually be
absorbed into the freedom to choose $c_0$ arbitrarily. 

Therefore, for $R^2$-gravity without matter, we have shown that, formulating the theory in terms of Feynman rules depending on only two coupling constants is self-consistent, provided the coupling constant associated with the Gauss-Bonnet term $G$ is correspondingly renormalized.  

We have only discussed the dimensionless coupling constants because, in a mass-independent renormalization scheme, the addition of UV irrelevant operators, such as an Einstein-Hilbert term or a cosmological constant, does not change the counterterms for the dimensionless couplings.  Thus, they may be added without consequences for this proof.  The preceding proof in no way required classical scale invariance.

This result may be rewritten in a number of other ways.  
Most commonly, $C^2$ is exchanged for $W$ as in \eqn{eq:l3}.  
We have that $\widetilde{c}=c+1/2a,$ so that 
$\beta_{\widetilde{c}}=\beta_c-\beta_a/(2a^2).$  
Therefore, to one-loop order, 
$\beta_{\widetilde{c}}=\kappa(-\beta_1+\beta_2/2),$
and the tree approximation to 
$\widetilde{c}$ will be
\beq
\widetilde{c}=c_0+\left(\half -\frac{\beta_1}{\beta_2}\right) \frac{1}{a}.
\eeq

\section{Extension to Matter}\label{sec:matter}

%\paragraph{ }
The results of the preceding section may be extended to the incorporation of matter with only slight modifications. 
The fundamental consistency relation must be extended to 
\beq\label{eq:consistency2}
\beta_{c} =\frac{\pa c}{\pa a}\beta_a
+\frac{\pa c}{\pa b}\beta_b+\sum_i\frac{\pa c}{\pa \lambda_i}\beta_{\lambda_i},
\eeq
where $\{\lambda_i\}$ represents all the additional dimensionless coupling constants in the theory.
 For example, the addition of a scalar field in the form
\beq\label{eq:jrealscalar}
S_m=\int d^4x \sqrt{g}\left[ \half (\nabla\phi)^2+\frac{\lambda}{4}\phi^4-\frac{\xi\phi^2}{2}R  \right].
\eeq
(Again, one could add mass terms or cubic couplings without changing the results for the dimensionless couplings.)
The divergences for $a$ and $c$ are modified by the inclusion of matter, but, at one-loop, they simply change the values of the constants $\beta_2$ and $\beta_1$, respectively~\cite{Fradkin:1981iu, Avramidi:1985ki}.   The nonminimal coupling $\xi$ adds to the divergences proportional to $R^2$, modifying the beta-function for $w$.  Similarly, the divergences for $\lambda$ as well as for the wave-function renormalization of $\phi$ receive gravitational contributions.  Their structure is well-understood~\cite{Buchbinder:1992rb}.  The form of these renormalizations can be brought into the same form as before as follows:  it turns out to be natural to rescale the field $\phi=\widetilde{\phi}/\sqrt{a}$ and coupling $\lambda\equiv a y$, so that the matter action 
\eqn{eq:jrealscalar} takes the form
\beq\label{eq:jrealscalar2}
S_m[\phi,g_\mn]=\!\int\! d^4x \frac{\sqrt{g}}{a} \left[ \half (\nabla\widetilde{\phi})^2+\frac{y}{4}{\widetilde{\phi}}^4-\frac{\xi{\widetilde{\phi}}^2}{2}R  \right]\!.\!
\eeq
Thus, $1/a$ factors out out, so that $a$ remains a loop-counting parameter, and the beta-functions for $w$, $y$, and $\xi$ may be written in the form
\beq
\frac{\pa w}{\pa u}=\overline{\beta}_w(w,\xi),\ 
\frac{\pa\xi}{\pa u}=\overline{\beta}_\xi(w,\xi,y),\  \frac{\pa y}{\pa u}=\overline{\beta}_y(w,\xi,y),
\eeq
where $du = -da/(\beta_2 a)$.  (See [{\bf I}] for further details.)
In this case, the fixed point behavior is far more complicated.  
We found that there are six fixed points, only one of which is a UV fixed point for all three couplings.  Its basin of attraction is limited and does not include all values of the couplings, which is to say that these parameters do not always approach finite fixed points.  It therefore depends on the initial conditions whether, as $a\to0$, all other couplings are AF or finite.  Thus, in the loop expansion of the renormalized couplings, the coefficients depend on the three parameters $w,\xi,y$ in general.

Finally, one may add scalars, fermions and non-Abelian gauge
fields.  Each species makes a contributions to the constants $\beta_1$
and $\beta_2$, but these gravitational couplings remain independent of
other coupling constants.  This is not true for $\beta_w,$ which can
depend on the nonminimal couplings $\xi_i$ of the scalar fields as well
as other dimensionless coupling constants. On the other hand, there are
more interrelated matter couplings that complicate the determination of
fixed points.  

We shall not discuss these in detail here, but some examples have been
worked out previously.  (For a summary of models, see 
Chapter~9 of \reference {Buchbinder:1992rb}.)
At one-loop, the gauge couplings receive
no contributions from the gravitational couplings, a vestige of their
classical conformal symmetry.  Generally, the Yukawa couplings
vanish more rapidly than the bosonic couplings, but the top quark coupling is so large in the SM that
it is often necessary to include it, at least up to the scale of the Planck mass, to obtain realistic predictions.
For present purposes, the important point is that none of these complications will alter the conclusions of this paper concerning the treatment of the couplings of the topologically-significant operators discussed herein.

\section{The Hirzebruch Signature}\label{sec:hirzebruch}

%\paragraph{ }
The G-B relation is not the only topologically motivated relation in theories such as these.  Another is the Hirzebruch signature whose topological density $R^*R$ is the gravitational contribution to the axial anomaly.
Our excuse for neglecting it until now is that, unlike $G$, it is only needed for renormalization in models that include fermions whose couplings imply $CP$-violation, as in the Standard Model.
Analogous to \eqn{eq:divb}, the local form of the relation 
\beq\label{eq:hirzebruch}
R^*R=\nabla_\mu H^\mu; \  H^\mu\!\equiv\! \epsilon^{\mn\gamma\delta}\Gamma^\rho_{\nu\kappa}\left[ \half R^\kappa{}_{\rho\gamma\delta}
+\frac{1}{3}\Gamma^\kappa_{\gamma\lambda}\Gamma^\lambda_{\rho\delta} \right]\!.
\eeq
Like $B^\mu,$ $H^\mu$ transforms like a connection.  The corresponding integral for for a compact manifold without boundaries is  
\beq\label{hirzebruchindex}
48\pi^2\tau=\int_\mcal\!d^4x\,\sqrt{g}\,R^*R=\int_\mcal d^4x \sqrt{g}\,C^*C,
\eeq
where the integer $\tau$ is referred to as the Hirzebruch signature or Hirzebruch index.  (If $\mcal$ has boundaries, then there will be additional terms representing their contributions.)
Since $(C\pm C^*)^2\ge0,$  $C^2\ge|C^*C|$, with equality only for $C=\pm C^*$ (self-dual or anti-self-dual.) Thus,
\beq
\int_\mcal\! d^4x\, C^2 \ge 48\pi^2|\tau|.
\eeq 
Consequently, only a compact spacetime that is not conformally flat can have nonzero signature.
Since
 \beq
 C^*C=\left(\frac{C+C^*}{2}\right)^2-\left(\frac{C-C^*}{2}\right)^2,
 \eeq
 it may come as no surprise that $\tau$ can be related to the number of self-dual $(b_2^+)$ or anti-self-dual $(b_2^-)$ harmonic two-forms.  In fact, $\tau=b_2^+ - b_2^-.$
 
The upshot of this is that another term may be added to the Lagrangian \eqn{eq:action1} of the form $i\vartheta C^*C.$  Obviously, $\vartheta$ is analogous to the $\theta$-parameter of QCD, and a nonzero value of $\vartheta$  implies the model is P- and CP-violating.  As with $G$, since $R^*R$ is a total derivative, $\vartheta$ will not contribute to the Feynman rules.  On the other hand, we expect that, if renormalized, it will obey an equation like \eqn{eq:consistency2}, since fermions contribute to the beta-functions for $a,$ $b$, as well as to those for other couplings, in particular, the beta-functions for Yukawa couplings.

\section{Instantons}\label{sec:instantons}

%\paragraph{}
The inclusion of these topologically significant terms in the action
suggests that they could become even more relevant nonperturbatively,
although this is not the focus of this paper.   There has been a great
deal of discussion about instantons in the context of Einstein-Hilbert
theory\footnote{For a review of early work, see, \eg,
\cite{Eguchi:1980jx}. \reference{Alvarez:1988tb} reviews some of the
subsequent developments.} and in string theory in higher dimensions,
especially their role in anomalies~\cite{AlvarezGaume:1983ig}.  For
higher-order gravity of the type considered herein, there has been
speculation about instantons and their potential effects assuming that
the theory has a sensible conformal limit~\cite{Strominger:1984zy,
Perry:1992ta}.  Although our work specifically assumes that the QFT is
not scale-invariant, let alone conformal-invariant, the potential
physical implications of instantons may well be similar to those that
were discussed for the conformal theory. Thus, an instanton that has a
nonzero Euler characteristic $\chi$ would presumably be
topology-changing, representing a tunneling amplitude from an initial
state that represents one genus (\eg, a sphere) to a final state that
represents another (\eg, a torus).  Similarly, if an instanton carries a
nonzero Hirzebruch signature $\tau,$ transitions between states of
different ``winding numbers'' should occur.  Since $\tau\ne0$ will
affect the chiral anomaly, it would be interesting to investigate what
changes, if any, such 
 instantons would imply for the usual picture of nonperturbative effects in QCD .  

We have not yet investigated the role of instantons in these theories,
but, as pointed out in \reference{Strominger:1984zy}, some of the
instantons presented there for the conformal theory ought to survive in
a scale-invariant theory, although these authors appear to have in mind
a theory without anomalies. Motivated by the considerations in this
paper and in [{\bf I}], we suggest that classically scale-invariant
theories may well provide a hospitable setting for treating such
instantons semi-classically, even though their QFT's {\bf are}
anomalous.   The point is that, at sufficiently high scales, their
background fields will be approximately scale-invariant By this,
we mean that, if all relevant couplings are AF, then the degree of
scale-breaking becomes small asymptotically.  Even if one supposes that
the background has constant curvature, the actual magnitude of the curvature
will still be undetermined. It remains to be seen whether topological
characteristics can be discussed within such a framework.
 
Earlier work assumes that instantons are ALE,  but in order to consider
spacetimes such as de~Sitter space, anti-de~Sitter space, and others
where curvature is essential and persistent, one must apparently give up
this requirement.  Exactly what alternative constraints are mandated for
such theories has yet to be determined.

For cosmological applications, one probably should be discussing only initial states with the time evolution determined by an ``in-in'' or Schwinger-Keldysh formalism\footnote{For some recent perspectives, see, \eg, Refs.~\cite{Adshead:2009cb, *Higuchi:2010xt, *Kaya:2012nn,*Kaya:2013nla}.}.  The Hartle-Hawking no-boundary hypothesis~\cite{Hartle:1983ai} is one such possibility, with a transition at the birth of the universe from Euclidean to Lorentzian signature.  It has been argued that such a framework strongly favors inflationary cosmologies~\cite{Hartle:2007gi,*Hartle:2008ng}.  
Just how starting from $R^2$ gravity and including instantons might affect such deliberations, if at all, remains unclear.

\section{Conclusions and Open Questions}\label{sec:conclusions}

We have demonstrated that, quite generally, renormalizable gravity
allows reduction from three to two primary operators and their
associated couplings, as required by the local Gauss-Bonnet relation in
four dimensions\footnote{Concerns about the viability of DREG were
expressed in \reference{Capper:1979pr}. One consequence of our results
is that these concerns have finally been laid to rest.  See also
\reference{deBerredoPeixoto:2004if}.}  This has been tacitly assumed
by previous authors, but there can be confusion concerning the precise
role of topological terms such as $G$ in the renormalization of the
theory, since it must be included among the renormalized operators. It
holds quite generally for the extension of pure gravity to include
matter consisting of an arbitrary collection of scalars, vectors, and
fermions.  A similar discussion undoubtedly applies to the Hirzebruch
signature density $C^*C,$ which is also a covariant divergence of a
``current'' $H^\mu.$  When fermions are added in such a way that CP is
violated, $\vartheta$ is expected to be renormalized, but in a manner
similar to $c(a,b).$  The idea then is that the only arbitrariness
in couplings such as $c(a,b)$ or $\vartheta(a,b)$ would be in the
constants $c_0$ and or $\vartheta_0.$  

One open question is whether the parameter $c_0,$
the only free parameter in the Gauss-Bonnet coupling, is in principle observable.  We
have our doubts that it can be observed in a purely perturbative
framework, but if instantons come to play a role in the determination of
acceptable states of the theory, then $c_0$ may well affect the
outcome. Similar remarks should apply to $\vartheta_0$ as well.

All these speculations presume that there are extensions of our earlier
work~[{\bf I}] to classically scale-invariant models in which there is
dimensional transmutation with an induced Planck mass in the same phase
in which the coupling constants are asymptotically free.  We suspect
that such models exist, as other authors have usually assumed about
models that explicitly break scale invariance
classically\footnote{A  recent application of this
type~\cite{Myrzakulov:2014hca} attempts  to include the effects of the
Gauss-Bonnet coupling.}

\begin{acknowledgments}
One of us (MBE) would like to thank G.~Horowitz for discussions concerning instantons.
DRTJ thanks Ian Jack for conversations, 
KITP (Santa Barbara), the Aspen Center for Physics and CERN for hospitality and
financial support, and the Baggs bequest for financial support. This research was supported in part by
the National Science Foundation under Grant No. NSF PHY11-25915 and
Grant No. PHYS-1066293, and by the Baggs bequest.  
\end{acknowledgments}

\begin{appendix}

\section{Background Field Quantization}\label{sec:bfq}

%\paragraph{}
In this section, we elaborate what we mean by the background field method of quantizing \eqn{eq:l4}.  This is completely standard, except for the way in which the term $c\,G$ enters the theory.  We shall follow the notation and conventions of Appendix~B of \reference{Einhorn:2014gfa}, employing DeWitt's condensed notation~\cite{DeWitt:1965jb}, using a single index to denote all indices, including spacetime $x^\mu$ or other continuous parameters.  Repeated indices are (usually) summed or integrated over.

For a classical action $S[\phi_i]$, the effective action may be formally defined by 
 $\Gamma[\phi_i]\equiv S[\phi_i]+\Delta\Gamma[\phi_i],$ where
\begin{subequations}
\begin{align}\label{eq:effaction}
e^{-\Delta\Gamma[\phi_i]}&=\int_\bcal \dcal h_i e^{-\Delta S[\phi_i,h_i]+h_k\frac{\delta \Delta\Gamma[\phi_i]}{\delta \phi_k}},\\
\label{eq:deltaclassical}
\mathrm{with}\ \Delta S[\phi_i,h_i]&\equiv
S[\phi_i+h_i]-S[\phi_i]-h_j\frac{\delta S[\phi_i]}{\delta \phi_j}.
\end{align}
\end{subequations}
$\bcal$ denotes the background manifold associated with $\phi_i.$
\eqn{eq:effaction} is a complicated integro-differential equation, whose meaning we have summarized previously in [{\bf I}].  Here, we want to focus on \eqn{eq:deltaclassical}, with $\phi_i$ replaced by the background metric, $g_\mn,$ and $h_i,$ by the metric fluctuations\footnote{An expression for $\Delta S$ to second order in $h_\mn$ may be found in \reference{Avramidi:2000pia}, eqns.~(4.53-4.55), spanning more than two full pages. To go beyond one-loop requires adding vertices arising in higher order. a formidable task!}, $h_\mn.$ The point is that, according to \eqns{eq:varindex}{eq:deltaclassical}, the operator $G$ enters only into the classical action $S[g_\mn]$ and not into $\Delta S[g_\mn,h_\mn]$, \eqn{eq:deltaclassical}, and therefore does not contribute to the integral \eqn{eq:effaction} that determines the QFT in the classical background.  The term $c\,G$ contributes neither to the propagator nor to the vertices.  

Next, since one is dealing with a gauge theory, one must add
gauge-fixing terms to $\Delta S,$ together with their associated
Faddeev-Popov ghosts, although for the most general background field,
this may not be necessary\footnote{The quadratic terms in $h_\mn$ may be
invertible without gauge-fixing, at least off-shell, which may be
sufficient for determining beta-functions.  For further discussion, see
the Appendix of \reference{Fradkin:1983mq}.}.  Of course, one then finds
that the Feynman rules lead to divergent integrals, so that the theory
must be regularized and renormalized.  The canonical procedure is to
express the classical action in terms of finite renormalized fields and
couplings plus divergent but local counterterms chosen to cancel these
divergences order-by-order in perturbation theory.  In the case of
interest, even though $G$ contributes nothing to the Feynman rules
arising from $\Delta S$, there are divergences arising that contribute
to the renormalization of the coupling $c$.  Such phenomena are
familiar already from QFT in curved spacetime even without quantizing the gravitational field.
(For example, see~\reference{Jack:1983sk}.)  If one is
to use DREG, this procedure requires extending the operators in the
classical action to $n$-dimensions.  This can easily be done for $C^2$,
$R^2$, as well as for $G$ in the form of  \eqn{eq:gbG} (but not in the
form of \eqn{eq:gb4}.)  

By this reasoning, we believe that there is no obstruction to renormalization (as there are with anomalies), and the renormalization program can proceed as usual.  Fortunately, we are not alone in our belief, inasmuch as this has also been implicitly assumed by all previous authors. 

Had one defined the QFT by extending the operators to $n$-dimensions at the outset, 
\eg, in the form of \eqns{eq:l1}{eq:l2}, or \eqn{eq:l3}, one could not use \eqn{eq:varindex} to develop the Feynman rules\footnote{In flat background, it has been shown by Zwiebach~\cite{Zwiebach:1985uq} that, even in higher dimensions, $G$ suprisingly remains a total derivative.  See also \reference{Zumino:1985dp}.}.  $\Delta S$ would include terms from $c\,G$ in the QFT contributing to the propagator and to vertices of order $\epsilon$ or higher.  In that case, it must be shown that the renormalized operators in four-dimensions actually respect \eqn{eq:varindex}, the Bianchi identities, and other special properties peculiar to the four-dimensional theory.  It would be nice to have a proof of this, but we
have not found such an argument in the literature.  Nevertheless, by our previous argument above, it seems that the coupling constant $c$ can be renormalized without including it in the Feynman rules for the QFT.

\section{Curvature in $n$-dimensions.} \label{sec:curvature}

The Riemann curvature $R^\kappa{}_{\mu\lambda\nu}$ can be defined in $n$-dimensions, from which one can obtain the Ricci tensor $R_\mn\equiv R^\lambda{}_{\mu\lambda\nu}$ and scalar $R\equiv R^\mu{}_\mu.$  The Weyl tensor $C_{\kappa\lambda\mn}$ can then be defined by the linear relation\footnote{Our convention concerning bracketed indices is to antisymmetrize those contained, so, \eg,
$g_{\kappa[\mu}{\widehat R}_{\nu]\lambda}\!\equiv\! g_{\kappa\mu}{\widehat R}_{\nu\lambda}\!-\!g_{\kappa\nu}{\widehat R}_{\mu\lambda}.$  A common alternative takes the brackets to mean half the difference.}
\begin{align}
\begin{split}
 C_{\kl\mn}\!&\equiv\!R_{\kl\mn}\!-\frac{1}{n-2}\left(g_{\kappa[\mu}{\widehat R}_{\nu]\lambda}\!-\!g_{\lambda[\mu}{\widehat R}_{\nu]\kappa} \right)\!+\cr & \hskip10mm 
 -\frac{R}{n(n-1)}\left(g_{\kappa[\mu}g_{\nu]\lambda}\!\right)\!,
\end{split}
\end{align}
where ${\widehat R}_\mn\equiv R_\mn-g_\mn R/n.$  Exchanging the positions of the Riemann and Weyl tensors, we may regard this as the decomposition of the Riemann tensor into its irreducible components under $SO(n)$, symbolically as $R=C\oplus\widehat{R}\oplus R$.  
This decomposition is orthogonal in the sense that\hfill 
\vskip-7mm
\beq
\hskip15mm 
R_{\kl\mn}^2 = C_{\kl\mn}^2 +
\frac{4\widehat{R}_\mn^2}{n-2}+\frac{2R^2}{n(n-1)}.
\eeq
%\smallskip
\noindent In four-dimensions, this becomes
\beq
\hskip15mm R_{\kl\mn}^2=C_{\kl\mn}^2+2\widehat{R}_\mn^2+\frac{R^2}{6}.
\eeq
%\bigskip
\end{appendix}

%\newpage
%\bigskip

\end{document}